\def\~{{$\tilde{\phantom{a}}$}}

\documentclass [12pt] {article}
\usepackage{epsfig}
\usepackage{color}

\renewcommand{\arraystretch}{1.5}

\def\thebibliography#1{\section{References}\markboth
 {REFERENCES}{REFERENCES}\list
 {[\arabic{enumi}]}{\settowidth\labelwidth{[#1]}\leftmargin\labelwidth
 \advance\leftmargin\labelsep
 \usecounter{enumi}}
 \def\newblock{\hskip .11em plus .33em minus -.07em}
 \sloppy
 \sfcode`\.=1000\relax}
\def\upcite#1{\raise6pt\hbox{\scriptsize
\cite{#1}}}
  \def\lsim{\mathrel {\vcenter {\baselineskip 0pt \kern 0pt
    \hbox{$<$} \kern 0pt \hbox{$\sim$} }}}
    \def\gsim{\mathrel {\vcenter {\baselineskip 0pt \kern 0pt
    \hbox{$>$} \kern 0pt \hbox{$\sim$} }}}

\def\hline{\noalign{\hrule \vskip2pt}}

\def\|{\ifmmode\Vert\else \char`\|\fi}
\ifx\oldzeta\undefined                          
  \let\oldzeta=\zeta                            
  \def\zzeta{{\raise 2pt\hbox{$\oldzeta$}}}     
  \let\zeta=\zzeta                              
\fi

\ifx\oldchi\undefined                           
  \let\oldchi=\chi                              
  \def\cchi{{\raise 2pt\hbox{$\oldchi$}}}       
  \let\chi=\cchi                                
\fi

\def\frac#1#2{{#1 \over #2}}

\def\half{\ifinner {\scriptstyle {1 \over 2}}
   \else {1 \over 2} \fi}

\def\simge{\mathrel{%
   \rlap{\raise 0.511ex \hbox{$>$}}{\lower 0.511ex \hbox{$\sim$}}}}
\def\simle{\mathrel{
   \rlap{\raise 0.511ex \hbox{$<$}}{\lower 0.511ex \hbox{$\sim$}}}}

\def\buildchar#1#2#3{{\null\!                   
   \mathop#1\limits^{#2}_{#3}                   
   \!\null}}                                    
\def\overcirc#1{\buildchar{#1}{\circ}{}}

\def\slashchar#1{\setbox0=\hbox{$#1$}           
   \dimen0=\wd0                                 
   \setbox1=\hbox{/} \dimen1=\wd1               
   \ifdim\dimen0>\dimen1                        
      \rlap{\hbox to \dimen0{\hfil/\hfil}}      
      #1                                        
   \else                                        
      \rlap{\hbox to \dimen1{\hfil$#1$\hfil}}   
      /                                         
   \fi}                                         %

\def\subrightarrow#1{
  \setbox0=\hbox{
    $\displaystyle\mathop{}
    \limits_{#1}$}
  \dimen0=\wd0
  \advance \dimen0 by .5em
  \mathrel{
    \mathop{\hbox to \dimen0{\rightarrowfill}}
       \limits_{#1}}}                           

\def\overlay#1#2{\ifmmode%
\setbox0=\hbox{$#1$}%
\setbox1=\hbox to\wd0{\hss$#2$\hss}\else%
\setbox0=\hbox{#1}%
\setbox1=\hbox to\wd0{\hss#2\hss}\fi%
#1\hskip-\wd0\box1 }

\def\pmb#1{\leavevmode\setbox0=\hbox{#1}%
\kern-.02em\copy0\kern-\wd0
\kern.04em\copy0\kern-\wd0
\kern-.02em\raise.04em\box0 }

\def\vereq#1#2{\lower3pt\vbox{\baselineskip1.5pt \lineskip1.5pt
\ialign{$\m@th#1\hfill##\hfil$\crcr#2\crcr\sim\crcr}}}

\def\tensor#1{\protect\@ontopof{#1}{\leftrightarrow}{1.15}\mathord{\box2}}
\def\overstar#1{\protect\@ontopof{#1}{\ast}{1.15}\mathord{\box2}}
\def\overdots#1{\protect\@ontopof{#1}{\cdots}{1.0}\mathord{\box2}}
\def\overcirc#1{\protect\@ontopof{#1}{\circ}{1.2}\mathord{\box2}}
\def\loarrow#1{\protect\@ontopof{#1}{\leftarrow}{1.15}\mathord{\box2}}
\def\roarrow#1{\protect\@ontopof{#1}{\rightarrow}{1.15}\mathord{\box2}}

\def\@ontopof#1#2#3{%
{\mathchoice
{\@@ontopof{#1}{#2}{#3}\displaystyle\scriptstyle}%
{\@@ontopof{#1}{#2}{#3}\textstyle\scriptstyle}%
{\@@ontopof{#1}{#2}{#3}\scriptstyle\scriptscriptstyle}%
{\@@ontopof{#1}{#2}{#3}\scriptscriptstyle\scriptscriptstyle}%
}%
}

\def\@@ontopof#1#2#3#4#5{%
\setbox0=\hbox{$#4#1$}%
\setbox1=\hbox{$#5#2$}%
\setbox2=\hbox{}\ht2=\ht0 \dp2=\dp0 %
\ifdim\wd0>\wd1 %
\setbox1=\hbox to\wd0{\hss\box1\hss}%
\mathord{\rlap{\raise#3\ht0\box1}\box0}%
\else   %
\setbox1=\hbox to.9\wd1{\hss\box1\hss}%
\setbox0=\hbox to\wd1{\hss$#4\relax#1$\hss}%
\mathord{\rlap{\copy0}\raise#3\ht0\box1}%
\fi
}%

\def\lambdabar{\protect\@lambdabar}
\def\@lambdabar{%
\relax
\bgroup
\def\@tempa{\hbox{\raise.73\ht0
\hbox to0pt{\kern.25\wd0\vrule width.5\wd0
height.1pt depth.1pt\hss}\box0}}%
\mathchoice{\setbox0\hbox{$\displaystyle\lambda$}\@tempa}%
{\setbox0\hbox{$\textstyle\lambda$}\@tempa}%
{\setbox0\hbox{$\scriptstyle\lambda$}\@tempa}%
{\setbox0\hbox{$\scriptscriptstyle\lambda$}\@tempa}%
\egroup
}

\def\corresponds{{\lower.2ex\hbox{=}}{\rm\kern-.75em^\triangle}}
\def\succsim{\succ\kern-.9em_\sim\kern.3em}
\def\precsim{\prec\kern-1em_\sim\kern.3em}
\def\slantfrac#1#2{\kern1em^{#1}\kern-.3em/\kern-.1em_{#2}}

\def\nhat{\widehat{\bf n}}

\begin{document}                                                                

\renewcommand{\arraystretch}{1.5}


\begin{center}
{\Large\bf Radiation from a Superluminal Source}
\\
\medskip

Kirk T.~McDonald \\

\smallskip
{\sl Joseph Henry Laboratories, Princeton University, Princeton, NJ 08544}
\\
(Nov.\ 26, 1986) \\

\medskip

{\bf Abstract}
\end{center}

The sweep speed of an electron beam across the face of an oscilloscope can
exceed the velocity of light, although of course the velocity of the electrons
does not.  Associated with this possibility there should be a kind of 
\v Cerenkov radiation, as if the oscilloscope trace were due to a charge
moving with superluminal velocity.

\section{Introduction}

The possibility of radiation from superluminal sources was first 
considered by Heaviside in 1888 \cite{Heaviside}.  He considered this topic
many times over the next 20 years, deriving most of the formalism of what is
now called \v Cerenkov radiation.  However, despite being an early proponent
of the concept of a velocity-dependent electromagnetic mass, Heaviside never
acknowledged the limitation that massive particles must have  velocities
less than that of light.  Consequently many of his pioneering efforts
(and those of his immediate followers, Des Coudres \cite{DesCoudres} and
Sommerfeld \cite{Sommerfeld}),
were largely ignored, and the realizable case of radiation from a charge
with velocity greater than the speed of light in a dielectric medium was
discovered independently in an experiment in 1934 \cite{Cerenkov34}.

In an insightful discussion of the theory of \v Cerenkov radiation, Tamm
\cite{Tamm39}
revealed its close connection with what is now called transition radiation,
{\it i.e.}, radiation emitted by a charge in uniform motion that crosses
a boundary between metallic or dielectric media.
The present paper was inspired by a work of Bolotovskii and Ginzburg
\cite{Bolotovskii} on how aggregates of particles can act to produce motion
that has superluminal aspects and that there should be corresponding
\v Cerenkov-like radiation in the case of charged particles.  The classic
example of aggregate superluminal motion is the velocity of the point of
intersection of a pair of scissors whose tips approach
one another at a velocity close to that of light.

Here we consider the example of a ``sweeping'' electron beam in
a high-speed analog oscilloscope such as the Tektronix 7104 \cite{Tek79}.
In this device the ``writing speed'', the velocity of the beam spot across
the faceplate of the oscilloscope, can exceed the speed of light.  
The transition radiation emitted by the beam electrons just before they
disappear into the faceplate has the character of \v Cerenkov radiation from
the superluminal beam spot, according to the inverse of the argument of
Tamm.

\section{Model Calculation}
 
As a simple model suppose a line of charge moves in the $-y$ direction with 
velocity $u \ll c$, where $c$ is the speed of light,
but has a slope such that the intercept with the $x$ axis moves with velocity
$v > c$.  See Figure 1a.
If the region $y < 0$ is occupied by, say, a metal the charges will 
emit transition radiation as they disappear into the metal's surface.
Interference among the radiation from the various charges then
leads to a strong peak in the radiation pattern at angle $\cos\theta = c/v$,
which is the \v Cerenkov effect of the superluminal source.

\begin{figure}[htp]  
\begin{center}
\includegraphics[width=3in, angle=0, clip]{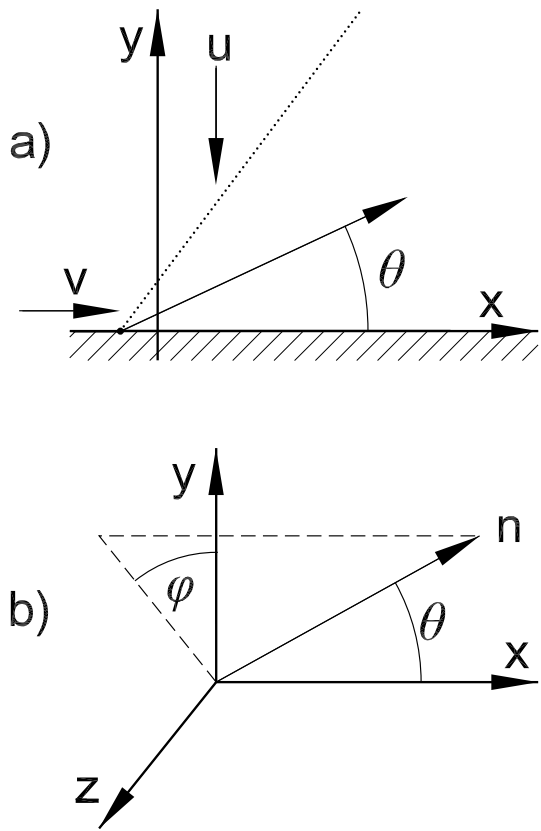}
\parbox{5.5in} 
{\caption[ Short caption for table of contents ]
{\label{ NAME } 
a) A sloping line of charge moves in the $-y$ direction with
velocity $v_y = u \ll c$ such that its intercept with the $x$ axis moves with
velocity $v_x = v > c$.  As the charge disappears into the conductor at 
$y < 0$ it
emits transition radiation.  The radiation appears to emanate from a
spot moving at superluminal velocity and is concentrated on
a cone of angle $\cos^{-1}(c/v)$.
 b) The angular distribution of the radiation is discussed
in a spherical coordinates system about the $x$ axis.
}}
\end{center}
\end{figure}

To calculate the radiation spectrum we use equation (14.70) from the textbook
of Jackson \cite{Jackson}:
\begin{equation}
{dU \over d\omega d\Omega} = {\omega^2 \over 4\pi^2 c^3} \left[ \int dt\
d^3r\ {\nhat \bf \times j}({\bf r},t) e^{i\omega(t - (\nhat{\bf \cdot r})/c)} 
\right]^2,
\label{eq1}
\end{equation}
where $dU$ is the radiated energy in angular frequency interval $d\omega$
emitting into solid angle $d\Omega$,
${\bf j}$ is the source current density, and $\nhat$ is a unit vector
towards the observer.

The line of charge has equation
\begin{equation}
y = {u \over v}x - ut, \qquad z = 0,
\label{eq2}
\end{equation}
 so the current density is
\begin{equation}
{\bf j} = -{\bf \widehat y}Ne \delta(z) \delta\left(t - {x \over v} + 
{y \over u} \right),
\label{eq3}
\end{equation}
where $N$ is the number of electrons per unit length intercepting the $x$ axis, 
and $e < 0$ is the electron's charge.

We also consider the effect of the image
current,
\begin{equation}
{\bf j}_{\rm image} = +{\bf \widehat y}(-Ne) \delta(z) \delta\left(t - 
{x \over v} - {y \over u}\right).
\label{eq4}
\end{equation}
We will find that to a good approximation the image current just
doubles the amplitude of the radiation.  
For $u \sim c$ the image current would be related to the
retarded fields of the electron beam, but we avoid this complication when
$u \ll c$.
Note that the true current exists only for $y > 0$, while the image current
applies only for $y < 0$.

We integrate using rectangular coordinates, with components of the unit vector
${\bf n}$ given by
\begin{equation}
n_x = \cos\theta, \qquad n_y = \sin\theta\cos\phi, \qquad{\rm and\ } \qquad
n_z = \sin\theta\sin\phi,
\label{eq5}
\end{equation}
as indicated in Fig.~1b.
The current impinges only on a length $L$ along the $x$ axis.
The integrals are elementary and we find, noting $\omega/c = 2\pi/\lambda$,
\begin{equation}
{dU \over d\omega d\Omega} = {e^2N^2L^2 \over \pi^2c} {u^2 \over c^2}
{\cos^2\theta + \sin^2\theta\sin^2\phi \over \big(1 - {u^2 \over c^2} 
\sin^2\theta\cos^2\phi \big)^2} 
\left({\sin\left[{\pi L \over \lambda}\big({c \over v} - \cos\theta\big)\right]
\over {\pi L \over \lambda}\big({c \over v} - \cos\theta\big)} \right)^2.
\label{eq6}
\end{equation}
The factor of form $\sin^2\chi/\chi^2$ appears from the $x$ integration, and
indicates that this leads to a single-slit interference pattern.

We will only consider the case that $u \ll c$, so from now on we approximate the
factor $1 - {u^2 \over c^2} \sin^2\theta\cos^2\phi$ by 1.

Upon integration over the azimuthal angle $\phi$ from $-\pi /2$ to
$\pi /2$ the factor $\cos^2\theta + \sin^2\theta\sin^2\phi$ becomes 
${\pi \over 2}(1 + \cos^2\theta)$.

It is instructive to replace the radiated energy by the number of radiated
photons: $dU = \hbar\omega dN_\omega$.  Thus
\begin{equation}
{dN_\omega \over d\cos\theta} = {\alpha \over 2\pi} {d\omega \over \omega}
N^2L^2 {u^2 \over c^2} (1 + \cos^2\theta)
\left({\sin\left[{\pi L \over \lambda}\big({c \over v} - \cos\theta\big)\right]
\over {\pi L \over \lambda}\big({c \over v} - \cos\theta\big)} \right)^2,
\label{eq7}
\end{equation}
where $\alpha = e^2/\hbar c \approx 1/137$.
This result applies whether $v < c$ or $v > c$.  But for $v < c$, the
argument $\chi = {\pi L \over \lambda}\big({c \over v} - \cos\theta\big)$ can
never become zero, and the diffraction pattern never achieves a principal
maximum.  The radiation pattern remains a slightly skewed type of transition
radiation.  However, for $v > c$ we can have $\chi = 0$, and the radiation
pattern has a large spike at angle $\theta_{\rm \check C}$ such that
$$ \cos\theta_{\rm \check C} = {c \over v},$$
which we identify with \v Cerenkov radiation.  Of course the side lobes are
still present, but not very prominent. 

\section{Discussion}

The present analysis suggests that \v Cerenkov radiation is not really distinct
from transition radiation, but is rather a special feature of the transition
radiation pattern which emerges under certain circumstances.  This viewpoint
actually is relevant to \v Cerenkov radiation in any real device which has
a finite path length for the radiating charge.  The walls which define the
path length are sources of transition radiation which is always present even
when the \v Cerenkov condition is not satisfied.  When the \v Cerenkov 
condition is satisfied, the so-called formation length for transition
radiation becomes longer than the device, and the \v Cerenkov radiation can
be thought of as an interference effect.

If $L/\lambda \gg 1$, then the radiation pattern is very
sharply peaked about the \v Cerenkov angle, and we may integrate over $\theta$
noting
\begin{equation}
d\chi = {\pi L \over \lambda} d\cos\theta \qquad {\rm and} \qquad
\int^\infty_{-\infty} d\chi {\sin^2\chi \over \chi^2} = \pi
\label{eq8}
\end{equation}
to find
\begin{equation}
dN_\omega \sim {\alpha \over 2\pi} (N\lambda)^2 {d\omega \over \omega}
{L \over \lambda} {u^2 \over c^2} \left(1 + {c^2 \over v^2}\right).
\label{eq9}
\end{equation}
In this we have replaced $\cos^2\theta$ by $c^2/v^2$ in the vicinity of the
\v Cerenkov angle.  We have also extended the limits of integration on $\chi$
to $[-\infty,\infty]$.  This is not a good approximation for $v < c$, in which 
case $\chi > 0$ always and $dN_\omega$ is much less than stated.  For $v = c$ 
the radiation rate is still about one half of the above estimate.

For comparison, the expression for the number of photons radiated in the
ordinary \v Cerenkov effect is
\begin{equation}
dN_\omega \sim 2\pi\alpha {d\omega \over \omega} {L \over \lambda}
\sin^2\theta_{\rm \check C}.
\label{eq10}
\end{equation}
The ordinary \v Cerenkov effect vanishes as $\theta^2_{\rm \check C}$ near the
threshold, but the superluminal effect does not.  This is related to the fact
that at threshold ordinary \v Cerenkov radiation is emitted at small angles to
the electron's direction, while in the superluminal case the radiation is at
right angles to the electron's motion.  In this respect the moving spot on an
oscilloscope is not
fully equivalent to a single charge as the source of the \v Cerenkov radiation.

In the discussion thus far we have assumed that the electron beam is well
described by a uniform line of charge.  In practice the beam is discrete,
with fluctuations in the spacing and energy of the electrons.  If
these fluctuations are too large we cannot expect the transition radiation
from the various electrons to superimpose coherently to produce the \v Cerenkov
radiation.  Roughly, there will be almost no coherence for wavelengths
smaller than the actual spot size of the electron beam at the metal surface,
Thus there will be a cutoff at high frequencies which serves to limit the
total radiated energy to a finite amount, whereas the expression derived
above is formally divergent.  Similarly the effect will be quite weak unless
the beam current is large enough that $N\lambda \gg 1$.

We close with a numerical example inspired by possible experiment. A realistic
spot size for the beam is 0.3 mm, so we must detect radiation at longer
wavelengths.  A convenient choice is $\lambda = 3$ mm, for which commercial
microwave receivers exist.  The bandwidth of a candidate receiver is 
$d\omega/\omega =
0.02$ centered at 88 GHz.  We take $L = 3$ cm, so $L/\lambda = 10$ and the
\v Cerenkov `cone' will actually be about $5^\circ$ wide, which happens to
match the angular resolution of the microwave receiver.  Supposing the
electron beam energy to be 2.5 keV, we would have $u^2/c^2 = 0.01$.  The
velocity of the moving spot is taken as $v = 1.33c = 4 \times 10^{10}$ cm/sec, 
so the observation angle is $41^\circ$.  If the electron beam current is 1
$\mu$A then the number of electrons deposited per cm along the metal surface
is $N \sim 150$, and $N\lambda \sim 45$.  

Inserting these parameters into the
rate formula we expect about $7 \times 10^{-3}$ detected photons from a single 
sweep of the electron beam.  This supposes we
can collect over all azimuth $\phi$ which would require some suitable optics.
The electron beam will actually be swept at about 1 GHz, so we can collect
about $7 \times 10^6$ photons per second.  The corresponding signal power
is $2.6 \times 10^{-25}$ Watts/Hz, whose equivalent noise temperature is
about 20 mK. This must be distinguished from
the background of thermal radiation, the main source of which is in the
receiver itself, whose noise temperature is about 100$^\circ$K
\cite{Wilkinson}.  A lock-in
amplifier could be used to extract the weak periodic signal; an integration
time of a few minutes of the 1-GHz-repetition-rate signal would suffice 
assuming 100\% collection efficiency.

Realization of such an experiment with a Tektronix 7104 oscilloscope would
require a custom cathode ray tube that permits collection of microwave
radiation through a portion of the wall not coated with the usual
metallic shielding layer \cite{Stoneman}.

\section{Appendix: Bremsstrahlung}

Early reports of observation of transition radiation were considered by
sceptics to be due to bremsstrahlung instead.  The distinction in principle
is that transition radiation is due to acceleration of charges in a medium
in response to the far field of a uniformly moving charge, while
bremsstrahlung is due to the acceleration of the moving charge in the
near field of atomic nuclei.  In practice both effects exist and can be
separated by careful experiment.

Is bremsstrahlung stronger than transition radiation in the example
considered here?  As shown below the answer is no, but even if it were we would 
then expect a \v Cerenkov-like
effect arising from the coherent bremsstrahlung of the electron beam as it
hits the oscilloscope faceplate.

The angular distribution of bremsstrahlung from a nonrelativistic electron
will be $\sin^2\theta$ with $\theta$ defined with respect to the direction
of motion.
The range of a 2.5-kev electron in, say, copper is about $5 \times 10^{-6}$
cm \cite{PB-212} while the skin depth at 88 GHz is about $2.5 \times 10^{-5}$
cm.  Hence the copper is essentially transparent to the backward hemisphere
of bremsstrahlung radiation, which will emerge into the same half space
as the transition radiation.

The amount of bremsstrahlung energy $dU_B$ emitted into energy interval
$dU$ is just $YdU$ where $Y$ is the so-called  bremsstrahlung yield factor.
For 2.5-keV electrons in copper, $Y = 3 \times 10^{-4}$ \cite{PB-212}.
The number $dN$ of bremsstrahlung photons of energy $\hbar\omega$ in a bandwidth
$d\omega/\omega$ is then $dN = dU_B/\hbar\omega = Yd\omega/\omega$.  For the
2\% bandwidth of our example, $dN = 6 \times 10^{-6}$ per beam electron.
For a 3-cm-long target region there will be 500 beam electrons per sweep
of the oscilloscope, for a total of $3 \times 10^{-4}$ bremsstrahlung photons
into a 2\% bandwidth about 88 GHz.  Half of these emerge from the faceplate as
a background to $7 \times 10^{-3}$ transition-radiation photons per sweep.  
Altogether, the bremsstrahlung
contribution would be about 1/50 of the transition-radiation signal in the
proposed experiment.

\end{document}